\documentclass[vecphys]{svmult}

\usepackage{makeidx}         
\usepackage{graphicx}        
\usepackage{multicol}        
\usepackage{cite}            
\usepackage[bottom]{footmisc}

\usepackage{color}
\newsavebox{\mysquare}
\savebox{\mysquare}{\textcolor{red}{\rule{0.1truecm}{0.1truecm}}}

\makeindex             

\def\tJ{t\textrm{-}J}
\def\tJV{t\textrm{-}J\textrm{-}V}
\def\hc{\mathrm{h.c.}}

\def\up{\uparrow}
\def\down{\downarrow}

\def\s{\sigma}

\begin{document}

\title{Mobile holes in frustrated quantum magnets and itinerant fermions
on frustrated geometries}
\titlerunning{Doped frustrated magnets}
\author{Didier Poilblanc \and Hirokazu Tsunetsugu}
\institute{Laboratoire de Physique Th\'eorique,
C.N.R.S. and Universit\'e de Toulouse III, 118, route de Narbonne,
31062 Toulouse cedex, France.
\texttt{didier.poilblanc@irsamc.ups-tlse.fr}
\and 
Institute for Solid State Physics, 
University of Tokyo, 
Kashiwanoha 5-1-5, Chiba 277-8581, Japan. 
\texttt{tsune@issp.u-tokyo.ac.jp}
}

\maketitle

\vskip 0.5truecm

\section{Introduction}
\label{ssec:1}

As discussed in several chapters
of this volume, frustration leads to unconventional (insulating) ground states.  
On the other hand doped holes are 
known to have profound effects in Mott insulators. Therefore doped frustrated 
systems offer the prospect of novel phases with some of the most fascinating, 
challenging and exotic behaviour. In addition, at commensurate electron fillings and in the presence of strong (screened)
Coulomb repulsion, 
geometrical frustration can also manifests itself as an extensive degeneracy of the classical ground-state manifold providing profound similarities with 
the field of quantum frustrated magnetism. 

Magnetic frustration in quantum spin systems leads frequently to the 
formation of spin singlets (dimers). Generically, systems of fluctuating quantum dimers can 
often order, breaking lattice symmetries to give rise to Valence Bond 
Crystals (VBCs)~\cite{VBC}, but under other circumstances they may 
remain in a quite unconventional quantum disordered state, the spin 
liquid, which breaks neither spin nor lattice symmetries. Anderson's 
original $d$-wave Resonating Valence Bond (RVB) state~\cite{RVB} is a 
paradigm for the spin liquid (in fact, for a particular type of gapless 
spin liquid, while the short-range RVB state composed of only nearest-neighbor 
dimers is gapped spin liquid). In a number of cases, frustrated spin systems 
and/or dimer systems can be doped, for example by chemical substitution in
a Mott insulator. When both spin and charge degrees of freedom are present, 
the role of frustration becomes unclear, and to date remains only poorly 
explored. It is, however, clear that new and exotic phenomena emerge upon 
doping, including heavy-fermion behavior, spin-charge separation or 
quasiparticle fractionalization, unconventional superconductivity, stripe
formation, bond and/or charge ordering, and many others. Such fundamental 
issues have motivated an increasing number of recent investigations, as 
well as the continuing search for new, doped materials. 

In this chapter, we describe some selected topics which illustrate the 
richness and diversity of the field of doped, frustrated magnets. The first
example concerns the dynamics of a small number of doped holes in the 
two-dimensional (2D) kagome and checkerboard Heisenberg quantum 
antiferromagnets. Without doping, the kagome Heisenberg antiferromagnet 
is believed to be a serious candidate for spin-liquid behavior, while 
the checkerboard lattice is the 2D analog of the well-known and highly 
frustrated 3D pyrochlore structure common in real materials. With doping, 
issues such as particle fractionalization and pairing can be addressed. In a 
second example, we move to the topic of Quantum Dimer Models (QDMs), similar 
to those proposed by Rokhsar and Kivelson in the context of the pseudo-gap
phase of high-temperature superconductors. Two classes of (weakly) doped 
QDMs will be discussed, which differ in the assumed statistics, bosonic or 
fermionic, of the bare holes. We proceed further by considering strongly 
correlated electrons on frustrated triangular lattices, and discuss the 
physics of an unconventional, reentrant metal-insulator transition.
As our final example, we consider correlated fermions moving on frustrated 
lattices at special, commensurate densities for which exotic but once again 
insulating ground states (GSs) are stabilized. For systems with strong 
interactions (Mott insulators), we show briefly how their behavior is 
analogous to the physics of (undoped) QDMs.

\section{Doping holes in frustrated quantum magnets}
\label{sec:2}

\subsection{The holon-spinon deconfinement scenario}
\label{ssec:20}

We begin our discussion of the phenomena associated with doped holes 
in frustrated magnets by considering the most popular paradigm for a 
nonmagnetic quantum ground state, which is a dimer-based system. If we 
assume that a single hole is ``injected,'' then the removal of the 
electron results in the breaking of one of the dimers, leaving behind
an empty site (holon) and a free spin (spinon) on the same bond. 

\begin{figure}[htbp]
  \centerline{\includegraphics*[width=0.5\linewidth]{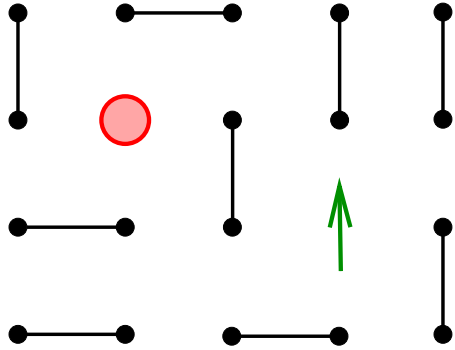}}
  \caption{\label{fig:Deconfinement}
Schematic representation of a holon (empty site, red) and a spinon 
(free spin, green) embedded in a fluctuating background of bond 
singlets (black). }
\end{figure}

If the dimers can change their positions by quantum fluctuations, the holon 
and the spinon can move on the lattice, across the diagonals of the 
plaquettes in a square-lattice system, by exchanging with the dimers. 
A typical configuration is shown in Fig.~\ref{fig:Deconfinement}. 
Optimization of their kinetic energies would require the holon and spin 
to delocalize, and thus to become separated. In a (gapped) spin liquid, 
realized if the system can fluctuate through all possible dimer coverings, 
a complete separation, known as ``deconfinement,'' is possible. In this
situation, the Landau quasiparticle (QP) breaks apart into separate species,
and an experimental technique which probes the hole Green function, such as 
Angle-Resolved Photoemission Spectroscopy (ARPES), would then show a broad 
maximum in place of the sharp QP peak. However, if dimer VBC order is present, 
{\it i.e.}~only one specific (type of) covering lies lowest in energy, one 
expects an effective string potential that binds the holon and spinon: 
indeed, if the dimers had no internal structure, an attempt to separate 
these two ``particles'' would lead to a continuous and linear increase of 
their energy. In reality, this increase is bounded by the spin gap (the 
energy to break up a dimer), beyond which which pairs of spinons would 
be generated spontaneously along the string.

\subsection{Single hole doped in frustrated Mott insulators}
\label{ssec:2a}

\begin{figure}[htpb]
  \centerline{\includegraphics*[width=0.7\linewidth]{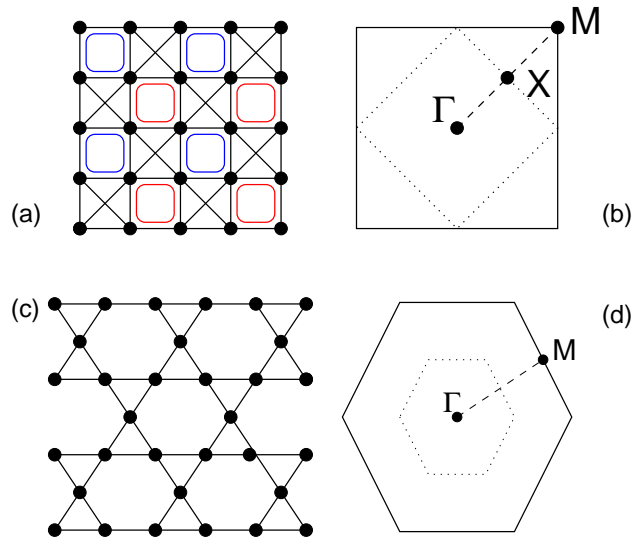}}
  \caption{\label{fig:Lattices}
(a) Checkerboard lattice. The two degenerate VBC GSs of the 
    nearest-neighbor Heisenberg model (half-filling) are represented 
    schematically in blue and red. This figure also serves as a 
    representation of the RSPC GSs of the quarter-filled, large-$V$ 
    Hubbard model of Subsection~\protect\ref{ssec:3b}.
(b) Full Brillouin Zone (BZ) of the checkerboard lattice.  The dotted 
    line corresponds to the reduced BZ associated with the $\sqrt{2} \!
    \times \! \sqrt{2}$ super-cell reflection. The path from the zone 
    center ($\Gamma$) to the ${\bf k}=(\pi,\pi)$ point ($M$) is shown
    as a dashed line. 
(c) Kagome lattice. 
(d) As in (b) but for the case of the kagome lattice. }
\end{figure}

Frustrated magnets are good candidates for the observation of spin-charge 
separation upon doping. The checkerboard and kagome lattices, shown in 
Figs.~\ref{fig:Lattices}(a) and (c), are good examples of the types of 
frustrated lattice on which such a phenomenon may be expected. They are 
composed, respectively, of strongly frustrated tetrahedra and triangles, 
assembled in a 2D, corner-sharing structure. While the AF Heisenberg 
Hamiltonian for $S = 1/2$ spins (in this chapter we consider only 
systems of $S = 1/2$ spins) on the checkerboard lattice is now 
thought to be a fully gapped system exhibiting VBC order (plaquette 
phase)~\cite{pyrochlore2D,Fouet,AltmanAuerbach,Brenig}, by contrast no 
sign of ordering has been found in the undoped kagome antiferromagnet 
\cite{KagomeED,KagomeED_2}, which possesses an exponentially large 
number of singlet states within the (finite-size) spin 
gap~\cite{KagomeED_2,KagomeSinglets}. It is these unconventional, 
low-lying excitations which open the door to new and surprising 
phenomena upon hole doping.

We perform exact diagonalization (ED) calculations based on the standard 
$t$-$J$ model Hamiltonian,
  \begin{equation}
    H_{\tJ}= - t \sum_{\langle i,j \rangle,\sigma}\ \mathcal{P} \left(
      c^{\dagger}_{i,\sigma} c_{j,\sigma} + \mbox{h.c.}\right) \mathcal{P}
    + J \sum_{\langle i,j\rangle}\ ({\bf S}_i \cdot {\bf S}_j - \frac{1}{4} 
    n_i n_j) ,
  \end{equation}
where on both lattices all bonds have the same couplings $t$, describing 
the kinetic energy of the hopping quasiparticles, and $J$, which is the 
superexchange interaction between the spins. This model is 
believed to be a reliable description of the low-energy physics of weakly 
doped Mott-Hubbard insulators with large optical excitation gaps. Here and 
hereafter, we assume the value $J = 0.4$ (in units where $|t|$ is set to 
$1$), which is the general order of the physical value in a number of real 
materials. The hole spectral functions are defined in the standard way as 
\begin{equation}
  A({\bf k},\omega) = -\frac{1}{\pi}
  \mbox{Im} \left[ \langle\Psi_0| c^\dagger_{{\bf k},\sigma} \frac{1}{\omega 
  + E_0 + i\eta - H} c_{{\bf k},\sigma} |\Psi_0 \rangle \right],
\end{equation}
and calculated by Lanczos ED, supplemented with a continued-fraction 
technique, on finite clusters with periodic boundary conditions to take 
advantage of the lattice translation symmetry. The reader is referred to 
the chapter of A. L\"auchli in this volume for a detailed discussion of 
numerical methods. Here we focus on the case of a single dynamic hole,
as studied in Ref.~\cite{Laeuchli_2004}. Because of the absence of 
particle-hole symmetry in frustrated lattices, it is necessary to 
distinguish between the cases $t > 0$ and $t < 0$. Note that for $t < 0$, 
frustration can also appear in the hole motion: in the example of a 
particle with the tight-binding dispersion on an isolated triangle, 
the kinetic-energy gain is $|t|$, a factor of two smaller than for $t > 0$. 

\begin{figure}[htpb]
    \centerline{\includegraphics*[width=0.7\linewidth]{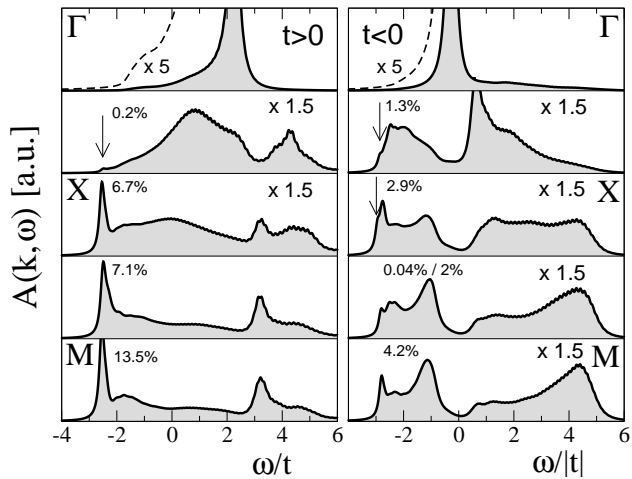}}
  \caption{\label{fig:SpectralCheckerboard}
    Single-hole spectral functions obtained on a 32-site checkerboard 
    cluster ($\sqrt{32} \! \times \! \sqrt{32}$, tilted at 45$^0$ to the 
    axes of Fig.~1.2) along the line $\Gamma M$. Left panel: 
    $t = +1$; right panel: $t = -1$. In both cases, $J/|t| = 0.4$. When 
    a quasiparticle peak is present, the corresponding weight is indicated. 
    Magnification factors applied in some cases are as indicated. From 
    Ref.~\protect\cite{Laeuchli_2004}.}
\end{figure}

Typical results obtained for a 32-site cluster on the checkerboard 
lattice are shown in Fig.~\ref{fig:SpectralCheckerboard} for the 
line $\Gamma M$ line in the Brillouin Zone (BZ) 
[Fig.~\ref{fig:Lattices}(b)]. At all points, most of the spectral weight 
is found to be incoherent, distributed over a range of 7-9$|t|$. However, 
a small QP peak is visible, particularly for momenta close to the $M$ point.
The region close to the $\Gamma$ point has only a very small QP peak, or 
possibly none at all, and the shape of the spectral function at $\Gamma$ 
itself is very special, probably because of its higher point-group symmetry.

\begin{figure}[htpb]
      \centerline{\includegraphics*[width=0.7\linewidth]{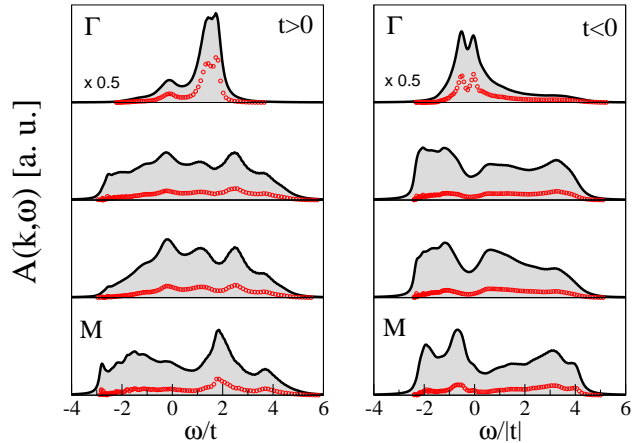}}
  \caption{\label{fig:Kagome27}
Single-hole spectral functions (black lines) along the line $\Gamma 
\leftrightarrow M$, computed on a 27-site kagome cluster for $t = +1$ 
(left panels) and $t = -1$ (right panels). Contributions of both singlet 
and triplet final states are included (see text). The red circles denote 
pole locations and their residues. From Ref.~\protect\cite{Laeuchli_2004}.}
\end{figure}

The analogous spectral functions of the kagome lattice, shown in
Fig.~\ref{fig:Kagome27}, show definite exotic behavior; they are very 
broad for all momenta (widths approximately 6-8$|t|$) and, in contrast 
to the checkerboard lattice, show no visible QP peaks, either for $t > 0$ 
(left panel) or for $t < 0$ (right panel). We stress that the broad
appearance of these spectra has no connection to the value of $\eta$ used 
in the calculation, but is an intrinsic feature of the spectral function, 
as can be deduced from the large number of poles carrying spectral weight 
(circles in Fig.~\ref{fig:Kagome27}). These spectral-function data support 
very strongly a spin-charge-separation scenario for the kagome lattice. 
Indeed, this spectacular phenomenon can be observed directly in the 
spin-density profile in the vicinity of the hole: a repulsion between 
the net $S = 1/2$ moment and the mobile hole is clearly visible, providing 
further support to the deconfinement scenario described above in the context 
of dimer-based systems. 

This investigation provides the first example of the observation of 
spin-charge separation in a 2D microscopic model. It establishes 
that the spin-liquid nature of the undoped ground state is crucial for such 
behavior. Indeed, in the checkerboard lattice, whose ground state exhibits 
a VBC structure, a weak holon-spinon confinement manifests itself as QP 
peaks in the spectral function for some momenta.

\subsection{Hole pairing and superconductivity}
\label{ssec:2b}

Whether doped holes could pair and lead to unconventional superconducting 
behavior is another of the fundamental issues raised recently by the new 
prospect of doping frustrated antiferromagnets. Indeed, superconductivity 
in the spinel oxide LiTi$_2$O$_4$~\cite{LiTi2O4}, in the recently 
synthesized 5$d$ transition-metal pyrochlores~\cite{5d-pyrochlore}, and 
in a layered triangular CoO compound~\cite{cobaltites} suggests that 
geometrical frustration, which could be magnetic and/or kinetic, might 
play a key role in the mechanism of unconventional superconductivity 
(as discussed in the chapter of Z. Hiroi and M. Ogata).

\begin{figure}[htpb]
\centerline{\includegraphics*[width=0.7\linewidth]{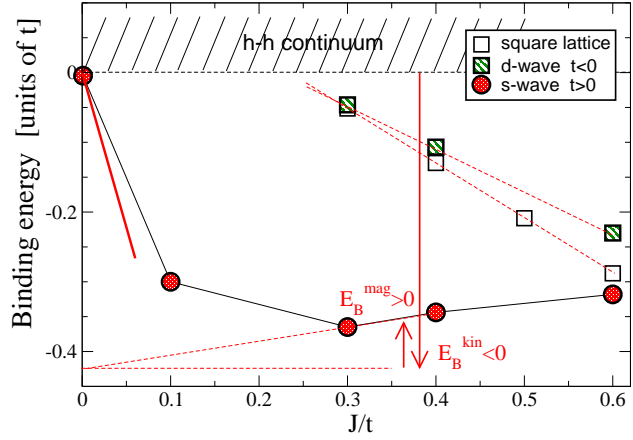}}
\caption{\label{fig:ener_2h}
Binding energies $E_B$ (in units of $|t| = 1$) of two holes doped on a 
square, 32-site cluster ($\sqrt{32} \! \times \! \sqrt{32}$) for $t > 0$ 
and $t < 0$, in the two orbital-symmetry channels corresponding to the 
respective GSs. The thick, red line corresponds the local slope at 
$J/t \rightarrow 0$. Red arrows show the large kinetic-energy gain and 
small loss in magnetic energy whose sum gives $E_B$. $E_B < 0$ is the 
signature of a bound state. From Ref.~\protect\cite{Poilblanc_2004}.}
\end{figure}

Cluster calculations (Fig.~\ref{fig:ener_2h}) were used to discover the 
occurrence of pairing in the doped checkerboard Heisenberg antiferromagnet 
described above~\cite{Poilblanc_2004}. It was shown that pairing, in several 
orbital channels including $s$- and $d$-wave, appears at arbitrarily small
$J/t$ for the particular sign of the hopping amplitude which leads to 
frustration in the motion of a single hole. In fact, hole delocalization 
({\it i.e.}~a gain in kinetic energy) plays a key role in this new mechanism 
of unconventional pairing, as also in some of the inter-layer tunneling 
mechanisms proposed by Anderson~\cite{Interlayer_PWA}. From these numerical 
data, a simple scenario might be proposed for $t > 0$: despite its suppressed
coherent motion, a single hole retains a strong incoherent motion, and thus 
can act to ``melt'' the plaquette VBC in its vicinity. This region, which 
may be somewhat extended in space, becomes more favorable for a second hole 
to gain kinetic energy, leading to correlated (or assisted) hopping. It is 
interesting to emphasize here the similarities with tight-binding studies 
of frustrated lattices, which show both localized single-particle states 
and interaction-induced, delocalized, two-particle bound 
states~\cite{correlated_hopping}.

In connection with cobaltates, superconductivity has also been investigated 
in the $t$-$J$ model on the triangular lattice, in particular using RVB 
variational Ans\"atze (presented in the chapter of Z. Hiroi and M. 
Ogata). In these studies, $d_{x^2-y^2} + id_{xy}$-wave superconductivity 
is found to be stable near half-filling. The relationship between this 
phase and the three-sublattice, 120-degree AF long-range order occuring 
at half-filling remains at present unclear and in need of further 
investigation.

\section{Doped Quantum Dimer Model}
\label{sec:4}

\subsection{Origin of the Quantum Dimer Model}
\label{ssec:4a}

The conclusions obtained in Sec.~\ref{ssec:2a} notwithstanding, both 
magnetic frustration and the introduction of fermionic variables (holes) 
lead, together or independently, to severe limits on the available numerical 
approaches. For example, quantum Monte Carlo algorithms, known to be very 
efficient for simple quantum spin systems, suffer from the infamous 
``minus-sign problem'' (introduced in the chapter of A. L\"auchli) and 
cannot be used at the required low temperatures. Practically, 
zero-temperature ED (by the Lanczos algorithm) and variational 
approaches are the only controlled methods practicable for systems 
such as the $t$-$J$ model on frustrated lattices. However, one alternative 
route to the investigation of microscopic models of the $t$-$J$ and Hubbard 
types is to construct effective models which would allow the use of more 
efficient methods, or calculations on larger clusters, while retaining the 
essential low-energy physics. 

\begin{figure}[htpb]
\begin{center}
\includegraphics[width=0.65\textwidth]{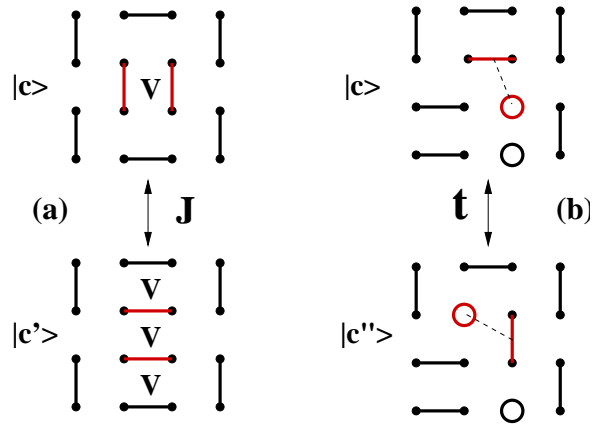}
\caption{ (a) Dimer-exchange process and (b) holon hopping process in 
the QDM.}
 \label{configs_QDM}
 \end{center}
 \end{figure}

When the effect of magnetic frustration is such that dimer degrees of 
freedom are relevant, one may consider the quantum hard-core dimer gas 
on a two-dimensional lattice. We illustrate this type of model by 
considering a square lattice, on which it is defined by the Hamiltonian
\begin{eqnarray}\label{QDM}
H_{\rm QDM} & = & V \sum_{c} N_c | c \rangle \langle c | -J \sum_{(c,c')} 
| c' \rangle \langle c | ,
\end{eqnarray}
where the sum over the index $(c)$ refers to all nearest-neighbor dimer 
coverings, $N_c$ is the number of ``exchangeable'' plaquettes, and the sum 
$(c',c)$ is over all pairs of configurations $| c \rangle$ and $| c' \rangle$ 
that differ by a single dimer-exchange process of the type illustrated in 
Fig.~\ref{configs_QDM}(a). In a manner similar to the square lattice, on 
the triangular lattice the exchange of parallel dimer pairs can be performed 
on the three different types of two-triangle rhombi. This model was introduced 
originally by Rokhsar and Kivelson~\cite{RK} in the context of the RVB 
theory of cuprate superconductors. The connection to the original spin 
formulation is, however, not completely clear: among other truncations 
of the spin degrees of freedom, the QDM of Eq.~(\ref{QDM}) deals by 
construction with orthogonal dimer coverings, which is not the case for 
the SU(2) dimer basis relevant in frustrated Heisenberg antiferromagnets. 
In spite of these subtleties, QDMs are expected to capture the essential 
physics of systems with singlet ground states, one primary reason for this 
being that they do possess the extreme ground-state degeneracy of the basis 
manifold. More details concerning these issues may be found in the chapter 
of R. Moessner in this volume. 

It is easy to introduce doping in the QDM. Holes may be injected only 
in pairs ({\it i.e.}~by removing dimers). However, doped holes can then
move independently by hopping between nearest-neighbor (on the triangular 
lattice) or diagonal next-nearest-neighbor (on the square lattice) 
sites~\cite{RK,doped_QDM,Ralko_2007_1}. The full Hamiltonian for a 
doped QDM is 
\begin{eqnarray}\label{doped_QDM}
H & = & H_{\rm QDM} - t \sum_{(c,c'')} | c'' \rangle \langle c | ,
\end{eqnarray}
where the $(c'',c)$ sum involves all pairs of configurations $| c \rangle$ 
and $| c''\rangle$, containing a fixed number $N_h$ of vacant sites (holes), 
that differ by a single hole hopping along a plaquette diagonal as illustrated 
in Fig.~\ref{configs_QDM}(b). In this formulation, bare holons, by which is 
meant the moving vacancies, have Bose statistics. Note that, in contrast to 
the triangular lattice, holes on the square lattice are constrained to 
remain only on one of the two sublattices. 

\begin{figure}[htpb]
\begin{center}
\includegraphics[width=0.65\textwidth]{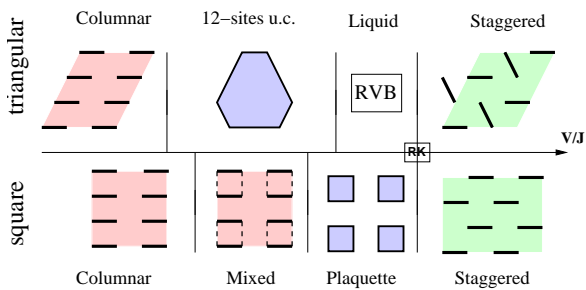}
\caption{Schematic phase diagrams of the undoped QDM for both the triangular 
(top) and square (bottom) lattices. Only the triangular lattice is believed 
to exhibit a liquid (RVB) phase. Evidence for the mixed phase on the square 
lattice has been reported in Ref.~\protect\cite{Ralko_2007_2}.}
 \label{ph_diag_QDM}
 \end{center}
 \end{figure}

\subsection{Phase diagrams at zero doping}
\label{ssec:4b}

Somewhat remarkably, the square- and triangular-lattice QDMs have quite 
different phase diagrams in the undoped case. First, exactly at $V = |J|$, 
which is known as the ``RK point,'' the GS takes the form of an equal 
superposition of all dimer coverings, and exhibits algebraic dimer 
correlations on the square lattice but short-ranged (exponentially 
decaying) correlations on the triangular lattice. Ordered VBC states 
appear on the square lattice immediately away from the RK point, whereas 
a gapped RVB liquid~\cite{moessner} is present over a finite region in 
$V/J$ on the triangular lattice. This RVB phase of the triangular lattice 
has also been shown to exhibit topological order~\cite{moessner,Ralko_Mila}, 
whose importance for frustrated systems is discussed in the chapter of G. 
Misguich, and for quantum information in the chapter of J. van den Brink, Z. 
Nussinov and A. M. Ole\'s. Comparative schematic phase diagrams for the two 
lattices are depicted in Fig.~\ref{ph_diag_QDM}. The square lattice shows a 
rich variety of VBC phases~\cite{syljuasen} with, in particular, a novel 
mixed phase~\cite{Ralko_2007_2} which interpolates between the columnar 
and the plaquette phases (the blue squares in Fig.~\ref{ph_diag_QDM} 
correspond to plaquettes on which vertical and horizontal dimer pairs 
resonate).

\subsection{Connection to the XXZ magnet on the checkerboard lattice}
\label{ssec:4c}

We have already explained the extent to which QDMs provide a natural 
framework to describe the dynamics of SU(2) singlets in frustrated but 
isotropic quantum antiferromagnets. We also mention briefly another case 
in which the QDM emerges as the model for the low-energy physics, that of 
the strongly anisotropic Heisenberg magnet (in the Ising limit) on the 
checkerboard lattice and in the presence of a magnetic field.

\begin{figure}
\begin{center}
\includegraphics[width=0.75\textwidth]{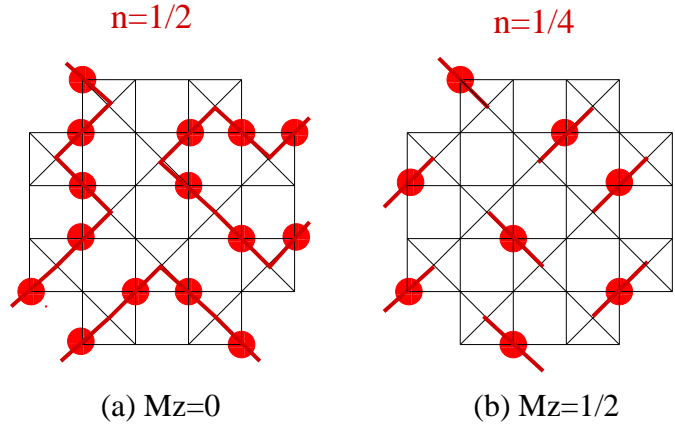}
\caption{ 
Representation of the Hilbert space of the large-$J_z$ XXZ Heisenberg model
on the checkerboard lattice in terms of (a) loop and (b) dimer coverings, 
corresponding respectively to (a) zero magnetization and (b) $M_z = \pm 
1/2$. The red dots represent hard-core bosons with, for example, $S_z = 1/2$ 
on the corresponding sites, while all other sites have $S_z = -1/2$. A dimer 
joining the centers of two ``tetrahedra'' is associated with each boson.
 \label{mapping_QDM} }
 \end{center}
 \end{figure}

We begin with no magnetic field and only an Ising coupling, $J_z S_i^z\, 
S_j^z$, on the bonds of the checkerboard lattice: in this case, the (classical) 
ground state is highly degenerate and can be fully represented by the ``loop 
coverings'' illustrated in Fig.~\ref{mapping_QDM}(a). Here, an up- (down-)spin
is represented by the presence (absence) of a dimer, or boson, on the bonds of 
an effective square lattice whose sites are in fact the centers of the 
``tetrahedra'' (the squares including diagonal bonds). The constrained 
nature of the classical GS is of the ``ice-rule'' type: the lowest Ising 
energy is obtained when there are precisely two bosons on every tetrahedron.
Second-order processes in the exchange coupling $J_{xy}$ lead to the dynamics 
of a six-vertex model~\cite{6VM} or a quantum loop model~\cite{QLM}.

By applying a magnetic field, the density of dimers (bosons) can be altered 
systematically. When an average of one dimer per tetrahedron is reached, 
again the ground-state manifold obeys an ice-rule constraint in the 
large-$J_z$ limit, where all states with precisely one dimer on every 
tetrahedron, as shown in Fig.~\ref{mapping_QDM}(b), are ground states. 
Second-order processes in $J_{xy}$ now lead to a QDM on the effective 
square lattice with $V = 0$ and $J = J_{xy}^2/J_z$ in Eq.~\ref{QDM}. 

\subsection{Bosonic doped Quantum Dimer Model}
\label{ssec:4d}

We turn now to the doped QDM and concentrate first on the case $J > 0$ 
in Eq.~\ref{doped_QDM}. In the mapping from SU(2) dimers, this sign of $J$ 
is expected for a  bosonic representation of the singlet bonds. The 
``monomers'' (holes) of the doped QDM discussed here could then be 
interpreted physically as entities such as unbound spinons. 

\vskip 0.4truecm
\begin{figure}
\begin{center}
\includegraphics[width=0.65\textwidth]{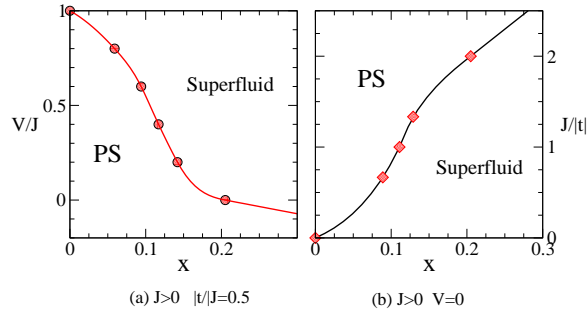}
\caption{ Schematic phase diagrams of the bosonic doped QDM as a function 
of dopant concentration $x$ and $V/J$ (a) or $J/|t|$ (b). Accurate GFMC 
data are obtained for $V/J \rightarrow 1$ and at moderate $|t|/J$ ratios 
(as explained in Ref.~\protect\cite{Ralko_2007_1}).
\label{phasediag_bosons} }
 \end{center}
 \end{figure}

For $J > 0$ and $t > 0$, the off-diagonal matrix elements of the Hamiltonian 
(\ref{doped_QDM}) are all non-positive, so that (from the Perron-Frobenius 
theorem) its GS has no node. Consequently, Green-function Monte Carlo (GFMC) 
techniques can be applied, particularly in the vicinity of the RK point 
and for small $t/J$ ratios ({\it i.e.}~when the exact RK GS is still a 
good guiding wave function), and the phase diagrams shown in 
Fig.~\ref{phasediag_bosons}(a,b) can be extracted by appropriate finite-size 
scaling. For larger values of $t/J$, such calculations can be complemented 
by ED on smaller clusters~\cite{Poilblanc_2007_1}. The phase-separation (PS) 
region consists of phase coexistence between an undoped VBC and a superfluid, 
the latter becoming stable as a unique component above a critical doping. 
It is notable that this superfluid exhibits flux quantization in units of
$h/2e$, in qualitative agreement with gauge theories of high-temperature 
superconductors~\cite{PALee} and recent, related $Z_2$ gauge 
theories~\cite{Z2_Fisher}.

\vskip 1.05truecm
\begin{figure}
\begin{center}
\includegraphics[width=0.65\textwidth]{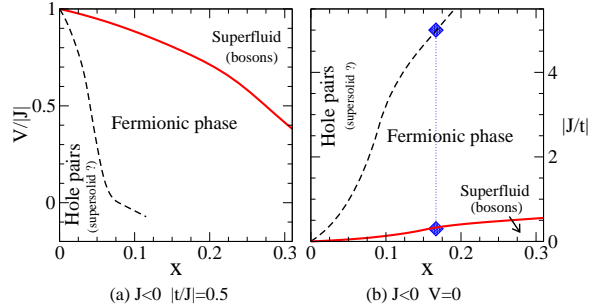}
\caption{ Schematic phase diagrams of the non-Frobenius ($J < 0$) doped 
QDM as a function of dopant concentration $x$ and $V/|J|$ (a) or $|J/t|$ 
(b), estimated by ED calculations~\protect\cite{Poilblanc_2007_1}.
\label{phasediag_fermions} }
\end{center}
\end{figure}
 
\subsection{Non-Frobenius doped Quantum Dimer Model on the square lattice}
\label{ssec:4e}

Turning to the case of Eq.~\ref{doped_QDM} with $J < 0$, a quite different 
type of behavior is expected. The ``non-Frobenius'' nature of the Hamiltonian 
(which prohibits the use of GFMC) reflects the original ``Fermi sign'' of 
the strongly correlated electrons. Indeed, if one interprets the dimers as 
SU(2) singlets, a dimer creation operator on the bond $ij$ can be written 
in the fermionic representation~\cite{Kivelson89,Vojta_Sachdev} as 
$d_{ij}^\dagger = (f_{i\uparrow}^\dagger f_{j\downarrow}^\dagger + 
f_{j\uparrow}^\dagger f_{i\downarrow}^\dagger)/\sqrt{2}$. In this basis, 
it can be verified that the effective dimer-exchange process generated by 
the underlying Heisenberg interaction within a plaquette occurs for the 
sign $J < 0$~\cite{Kivelson89}. In addition, the electron-destruction 
operator takes the form $c_{i\sigma} = f_{i\sigma} b_i^\dagger$, where 
the holon (hole or monomer) -creation operator $b_i^\dagger$ is bosonic.

The phase diagram of the non-Frobenius doped QDM obtained by 
ED~\cite{Poilblanc_2007_1} is both exotic and rich, as shown in
Fig.~\ref{phasediag_fermions}. First, bare holons can be seen binding to 
topological defects (namely vortices, also known as ``visons''), producing 
fermionic composite particles; alternatively stated, the hole becomes a 
fermion. Secondly, in contrast to the bosonic case, no PS is seen in the 
immediate vicinity of the (VBC) Mott insulator. Instead, a $d$-wave pairing 
is expected, opening the possibility of unconventional superconductivity. 
Finally, at large kinetic energies, holons and vortices unbind, bosonic 
holes Bose-condense, and a superfluid phase is stabilized, presumably of 
the same type (with $2e$ charge quanta) as that obtained for $J > 0$. 

\section{Mott transition on the triangular lattice}

\subsection{Frustration in itinerant electron systems}

In the first half of this chapter, we have discussed the dynamics of holes 
doped into magnetic insulators with geometrical frustration. Another 
important class of phenomena is driven by frustration in metallic systems.
Several strongly correlated metallic systems, such as LiV$_2$O$_4$ 
\cite{Kondo_1997} and (Y,Sc)Mn$_2$ \cite{Wada_1987}, show unusually large 
entropies at temperatures much lower than their bare energy scales (band 
width and Coulomb repulsion), and this is thought to be related to the 
geometrical frustration inherent in their lattice structure 
\cite{Pollmann,Frust_vanadate}. In strongly correlated electron systems, 
double occupancy of a site is suppressed by the large, on-site Coulomb 
repulsion, and the probability of single occupancy increases. This tends 
to enhance the formation of a local magnetic moment at each site, which 
interacts with neighboring moments, and frustrated configurations may be 
adopted depending on the lattice geometry. The central issues for frustrated 
metals are the effects of exotic magnetic fluctuations on quasiparticle 
coherence and novel magnetic long-range order or characteristic correlations 
in itinerant systems.  

\subsection{Mott transition in organic compounds with triangular geometry}

One of the well-studied problems in the physics of frustrated metals 
is the Mott metal-insulator transition on a triangular lattice 
\cite{Triangle_Imada,Parcollet_2004,Kyung_2006}. When the Coulomb repulsion 
is much larger than the band width ($U \gg W$), the half-filled system is 
described effectively by the Heisenberg spin model in the sector of energies 
below the Mott-Hubbard charge gap. In this case, the well-known 120$^\circ$ 
structure appears in the ground-state spin configuration. A more exotic 
situation can be expected when the Coulomb repulsion is comparable with 
the band width ($U \sim W$), when ring-exchange processes involving multiple 
sites become important. These processes, which are discussed in the chapter 
by G. Misguich, open the possibility of stabilizing exotic states. 

Experiments on organic compounds with a triangular lattice structure
\cite{Shimizu_2003,Kurosaki_2005,Kagawa_2004} have stimulated theoretical 
studies on the triangular-lattice Hubbard model. These materials are 
quasi-two-dimensional $\kappa$-(ET)$_2$$X$ systems with several possible 
monovalent anions $X$. ET denotes the bis(ethlylenedithio)-tetrathiafulvalene 
molecule, also often represented as BEDT-TTF, and dimerized pairs of ET 
molecules constitute a triangular lattice. Each pair provides one conduction 
electron, and the system is well described by a half-filled Hubbard model on 
the triangular lattice with nearest-neighbor hopping terms, 
\begin{equation}
H = \sum_{\langle i, j \rangle} \sum_{\sigma} 
t_{ij} c^\dagger_{i \sigma} c_{j \sigma} 
- \mu \sum_{i , \sigma} n_{i \sigma} 
+ U \sum_{i} n_{i \uparrow} n_{i \downarrow} . 
\end{equation}
Here $n_{i \sigma} = c^\dagger_{i \sigma} c_{i \sigma}$ and, because 
this model lacks electron-hole symmetry, the chemical potential $\mu$ is 
introduced to adjust the electron density to half filling, $\sum_\sigma 
\langle n_{i \sigma} \rangle = 1$. The organic ET compounds have in fact 
only intermediate correlation strengths: because each site in the model 
represents a pair of molecules and the corresponding Wannier wave function 
is extended over the size of this pair, the Coulomb repulsion $U$ is 
smaller than in the case of typical inorganic compounds, and as a result 
charge fluctuations have important effects. Because of the non-spherical shape 
of the molecule pairs, there are two different hopping integrals between 
nearest-neighbor sites, $t$ and $t'$ [Fig.~\ref{fig:PD_CDMFT}(a)].  
The ratio $t'/t$ depends on the anion 
species $X$, and is an important parameter controlling the frustration. 

\begin{figure}[tbh]
  \centerline{\includegraphics*[width=0.8\linewidth]{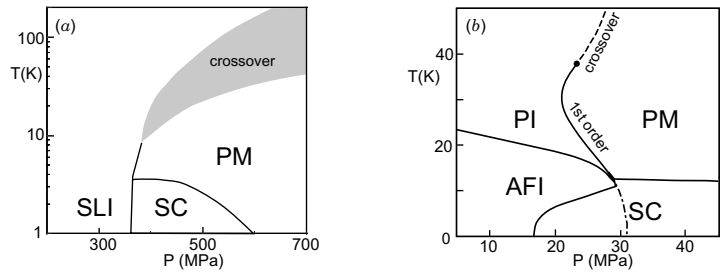}}
  \caption{\label{fig:PD_kappaET}
Temperature-pressure phase diagram of (a) $\kappa$-(ET)$_2$Cu$_2$CN$_3$ 
and (b) $\kappa$-(ET)$_2$Cu[N(CN)$_2$]Cl. PM: paramagnetic metal, PI: 
paramagnetic insulator, AFI: antiferromagnetic insulator, SLI: spin-liquid 
insulator, SC: superconducting phase. Reproduced based on 
Refs.~\cite{Kurosaki_2005} and \cite{Kagawa_2004}.   }
\end{figure}

Extensive investigation of the ET systems \cite{Shimizu_2003,Kurosaki_2005,
Kagawa_2004} has demonstrated that their low-energy magnetic properties 
change dramatically for different anions $X$. Particular highlights in 
the series include spin-liquid-like behavior in $\kappa$-(ET)$_2$Cu$_2$CN$_3$ 
\cite{Kurosaki_2005} and a reentrant metal-insulator transition with 
decreasing temperature at intermediate pressures in 
$\kappa$-(ET)$_2$Cu[N(CN)$_2$]Cl \cite{Kagawa_2004}. The difference between 
these systems lies in the different values of the frustration parameter 
$t'/t$. Quantum chemistry calculations estimate that $t'/t$ = 1.06 for 
$\kappa$-(ET)$_2$Cu$_2$CN$_3$ and 0.75 
for $\kappa$-(ET)$_2$Cu[N(CN)$_2$]Cl \cite{Shimizu_2003}.
From the viewpoint of their electronic 
structure, the candidate spin-liquid material $\kappa$-(ET)$_2$Cu$_2$CN$_3$ 
is very close to being a regular triangular system, in which all the hopping 
integrals are the same, while the reentrant material 
$\kappa$-(ET)$_2$Cu[N(CN)$_2$]Cl corresponds to a triangular geometry  
perturbed towards an unfrustrated square lattice.  

Figure \ref{fig:PD_kappaET} shows the temperature-pressure phase diagrams 
of these two compounds. In these systems, the primary effect of applying 
pressure is to increase the hopping integrals, and thus the region of higher 
pressure in experiments corresponds to smaller values of $U/W$ in the 
Hubbard model. The boundaries between the metallic and insulating 
phases in the two materials differ qualitatively in shape: in 
$\kappa$-(ET)$_2$Cu$_2$CN$_3$, the insulating phase appears on the 
high-temperature side of the boundary; in $\kappa$-(ET)$_2$Cu[N(CN)$_2$]Cl, 
this is only the case above approximately 30 K, while below this the 
insulating phase extends as the temperature decreases. The former type 
of behavior is beyond the naive expectation that electron dynamics become 
more coherent with decreasing temperature, but it is consistent with the 
conventional Mott transition, as we explain in detail in the next 
subsection.

\begin{figure}[bthp]
\centerline{\includegraphics*[width=0.9\linewidth]{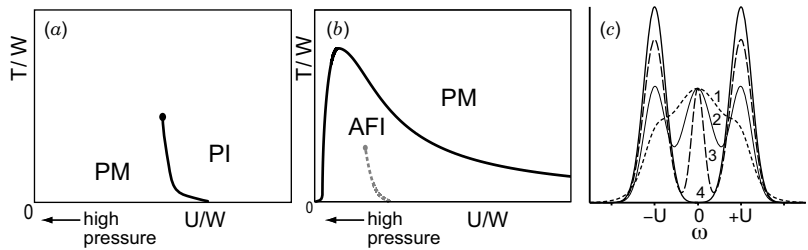}}
\caption{\label{fig:PD_DMFT} Phase diagram of the Hubbard model determined 
by dynamical mean-field theory: (a) single-site approximation and (b) 
results obtained when considering the possibility of AF long-range order, 
corresponding to an unfrustrated lattice. The dotted line marks the 
metal-insulator transition when magnetic order is absent. The same 
acronyms are used for labelling the phases as in Fig.~\ref{fig:PD_kappaET}.  
(c) Schematic illustration of the electron spectral function at the Mott 
transition, shown for different values of $U/W$ (1 $\rightarrow$ 4).}
\end{figure}

\subsection{Mott transition in the triangular-lattice Hubbard model}

A schematic phase diagram of the ``frustrated'' Hubbard model is shown in 
Fig.~\ref{fig:PD_DMFT}(a). There is a first-order phase transition separating 
metallic and insulating phases. As $U/W$ is increased, spectral weight is 
transferred from the region around $\omega = 0$ to the lower and upper Hubbard 
bands at $\omega \sim \pm U$ [Fig.~\ref{fig:PD_DMFT}(c)]. The central peak 
disappears at the transition point and the insulating phase is on the
high-temperature side of the boundary. The model is frustrated in the sense 
that these results are obtained from a single-site, dynamical mean-field 
theory \cite{DMFT}, which assumes the absence of a magnetic instability 
and thus describes well the case of strong frustration; the phase diagram 
shows a line of Mott transitions in the original sense of this term, 
meaning transitions occurring with no simultaneous magnetic order. In the 
insulating phase, each site has a finite static magnetic moment, which is 
effectively decoupled from the surrounding moments, and this leads to a 
large spin entropy of order $\log 2$. This is the reason that the insulating 
phase is stabilized at higher temperatures. In unfrustrated systems, as shown 
in Fig.~\ref{fig:PD_DMFT}(b), the metal-insulator transition takes place 
simultaneously with the emergence of AF long-range order, and the genuine 
Mott transition does not occur.

\begin{figure}[bthp]
\centerline{\includegraphics*[width=0.75\linewidth]{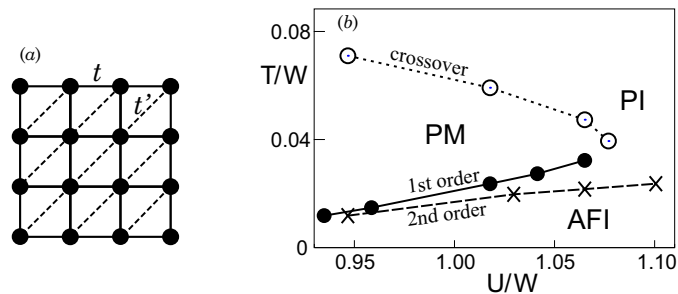}}
\caption{\label{fig:PD_CDMFT} (a) Generalized triangular lattice with two 
types of nearest-neighbor hopping integrals, $t$ and $t'$. The case $t = t'$ 
corresponds to the regular triangular lattice. (b) Phase diagram of the 
anisotropic, triangular lattice with $t' = 0.8t$.}
\end{figure}

We now return to the mysterious reentrant metal-insulator transition in 
$\kappa$-(ET)$_2$Cu[N(CN)$_2$]Cl. This problem was investigated in 
Ref.~ \cite{Ohashi_2008} by studying the half-filled Hubbard model 
on an anisotropic, triangular lattice [Fig.~\ref{fig:PD_CDMFT}(a)] with 
$t' = 0.8t$, to determine the $U$-$T$ phase diagram using cellular 
dynamical mean-field theory (CDMFT). This method is a generalization of
the conventional dynamical mean-field theory which uses a cluster of 
multiple sites (four for this system) \cite{CDMFT}, and allows one to 
calculate electronic Green functions and different correlation functions 
in addition to thermodynamic quantities at finite temperatures. The CDMFT 
approach has the advantage that both quantum and thermal fluctuations, and 
thus frustration effects, are taken into account completely inside the 
cluster.   

The phase diagram of the anisotropic, triangular-lattice Hubbard model is 
shown in Fig.~\ref{fig:PD_CDMFT}(b). The band width at $U = 0$, $W = 8.45t$ 
is taken as the unit of energy. The phase diagram was determined by analyzing 
the double occupancy $D \equiv \langle n_{i \uparrow} n_{i \downarrow}  
\rangle$, which is a measure of metallicity. In the metallic and insulating 
phases identified in this way, $D(T)$ decreases with decreasing temperature 
in the paramagnetic insulating (PI) phase, while it increases in the 
paramagnetic metallic (PM) phase. In the high-temperature regime, these 
two phases merge smoothly at the crossover line (dotted), which is defined 
by the condition $dD/dT = 0$.  In the low-temperature regime, the two phases 
are separated by a first-order transition, where the double occupancy shows 
an abrupt jump.  

\begin{figure}[bthp]
\centerline{\includegraphics*[width=\linewidth]{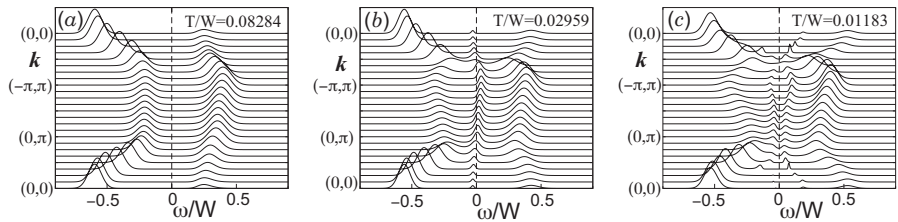}}
\caption{\label{fig:Hubbard_spectral} Wave vector-dependent electron 
spectral functions of the half-filled Hubbard model on an anisotropic, 
triangular lattice with $t' = 0.8t$ and $U/W = 0.947$.}
\end{figure}

It is to be noted that reentrant behavior of the metal-insulator 
transition/crossover is indeed found in the anisotropic, triangular-lattice 
Hubbard model for intermediate values, $t'/t \sim 0.8$, of the hopping 
ratio. The nature of this reentrant behavior is clearly visible in the 
wave vector-dependent electron spectral function $A_\mathbf{k} (\omega)$,
shown in Fig.~\ref{fig:Hubbard_spectral} for three representative 
temperatures and at the fixed value $U/W = 0.947$. In the high-temperature 
PI phase, there is a wide Hubbard gap in the spectrum around $\omega = 0$. 
In the intermediate PM phase, a heavy-quasiparticle band emerges in the 
Hubbard gap, a clear sign of metallic behavior and consistent with the 
conventional Mott transition depicted in Figs.~\ref{fig:PD_DMFT} (a) and 
(c). However, the low-temperature, first-order transition line has a 
different character: the heavy-quasiparticle band does not disappear, 
splitting instead into two bands separated by a small energy gap, as 
shown in Fig.~\ref{fig:Hubbard_spectral}(c). This behavior is similar 
to the case of a metal-insulator transition driven by magnetic instability.  

This type of explanation is confirmed by calculations of the magnetic 
susceptibility, $\chi_\mathbf{q}$ \cite{Ohashi_PTP}, which has a peak at 
the incommensurate wave vectors $\mathbf{Q} \approx \pm (0.7\pi,0.7\pi)$.  
This peak grows as the temperature decreases, and diverges at a finite 
temperature indicated by the crosses in the phase diagram of 
Fig.~\ref{fig:PD_CDMFT}(b). The line of magnetic instability is very 
close to the first-order metal-insulator-transition line, and it is 
reasonable to expect that the metal-insulator transition is driven by 
enhanced magnetic fluctuations. It should also be noted that the two 
lines are separate and there exists a finite region of a paramagnetic 
insulating phase between them.  

To summarize this section, the anisotropic, triangular-lattice Hubbard 
model has a phase diagram showing a reentrant metal-insulator transition.  
This phenomenon is a direct consequence of the effects of geometrical 
frustration on magnetic correlations. Taking increasing pressure to reduce 
the ratio $U/W$, the calculated phase diagram reproduces qualitatively the 
essential features of the phase diagram of $\kappa$-(ET)$_2$Cu[N(CN)$_2$]Cl.  
The finite values of the magnetic transition temperature in this 2D model 
are a consequence of the mean-field-type approximation made for 
inter-cluster correlations, but may provide an estimate of the true 
values which would be obtained on including the 3D couplings present 
in real materials. 

\section{Ordering phenomena at commensurate fermion densities on
frustrated geometries}
\label{sec:3}

In the preceding sections we have discussed only correlated systems at or 
near half-filling ({\it i.e.}~with one electron per lattice site). However, 
repulsive interactions with longer range than the on-site terms considered 
above can also give rise to insulating behavior at different commensurate 
densities. Examples include quarter-filling and even 1/8-filling on the 
checkerboard lattice, and we illustrate this phenomenon here by discussing 
two scenarios occuring on frustrated geometries. One is a Bond-Order-Wave 
(BOW) instability is driven directly by particular nesting properties of 
the Fermi surface. The other concerns the effects of nearest-neighbor 
interactions sufficiently strong that they produce a novel type of Mott 
insulator exhibiting an exotic VBC order. The properties of this Mott 
insulator may be described by an effective QDM, hence providing a formal 
connection with Sec.~\ref{sec:4}. This latter insulator can also be doped, 
a point we mention briefly as a possible route towards quite new and exotic 
metallic and superconducting behavior. 

\subsection{Bond Order Waves from nesting properties of the Fermi surface}
\label{ssec:3a}

Let us consider the extended Hubbard Hamiltonian, $H = H_0 + H_{\mathrm{int}}$,
on the 2D frustrated kagome and checkerboard lattices. We recall here that 
these lattices are composed of corner-sharing units (respectively triangles 
and tetrahedra) residing on an underlying bipartite lattice (respectively 
hexagonal and square), a point which will be important in determining their 
behavior. The kinetic part of the Hamiltonian is given by
\begin{equation}
\label{Hnull}
H_0 = - t\sum_{\langle ij \rangle} \sum_{\s = \up\down} \left( 
c^{\dagger}_{i\sigma} c^{\phantom\dagger}_{j\sigma} + \hc \right),
\end{equation}
with positive hopping matrix element $t$, and the sum $\sum_{\langle ij 
\rangle}$ is over all bonds on the lattice. The interaction part is given by 
\begin{equation}
\label{Hint}
 H_{\mathrm{int}} = U \sum_i n_{i\uparrow} n_{i\down} + J \sum_{\langle ij 
\rangle} \mathbf{S}_i \cdot \mathbf{S}_j + V \sum_{\langle ij \rangle} n_i 
n_j,
\end{equation}
with on-site repulsion $U$, nearest-neighbor spin exchange $J$, and 
nearest-neighbor repulsion $V$; in this subsection, we consider the 
regime of weak and intermediate couplings.

On kagome and checkerboard lattices, the non-interacting hamiltonian $H_0$ 
exhibits a dispersionless (flat) band which, for $t > 0$, lies at the
top of the spectrum and plays no role. For the kagome lattice, the band 
structure is remarkable for the presence of ``Dirac cones'' positioned 
exactly at the Fermi level of the 1/3-filled system ($n = 2/3$) and 
leading to semi-metallic behavior (also relevant on the 3D pyrochlore 
lattice). Renormalization-group and numerical techniques~\cite{Indergand1} 
have been applied to demonstrate that a spontaneous symmetry-breaking occurs 
for arbitrarily small interactions in this system: the instability corresponds 
to a BOW in which the kinetic energy is staggered for neighboring triangular 
units on the underlying hexagonal lattice. We stress that no charge modulation 
is present (so that all sites remain equivalent), the BOW breaking only the 
spatial site-inversion (180$^0$) symmetry, such that up- and down-pointing 
triangles in Fig.~\ref{fig:Lattices}(c) become different, while translational 
symmetry is preserved~\cite{Indergand1}.

\begin{figure}
\begin{center} 
\includegraphics[width=0.3\textwidth]{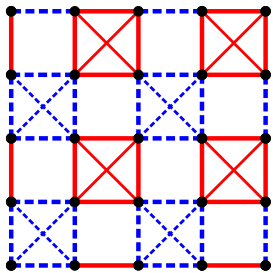}
\caption{Bond Order Wave on the checkerboard lattice at $n = 1/2$. The four
inequivalent bonds are represented by different colors/line types and by different thicknesses
(diagonal and x/y bonds are different). All the sites (black dots) carry the same electron occupancy (1/2 electron
on average).
\label{BOW} }
 \end{center}
 \end{figure}
 
On the checkerboard lattice, a BOW instability appears at quarter-filling ($n = 1/2$).
As on the kagome lattice, this BOW (Fig.~\ref{BOW}) is characterized by two 
types of (tetrahedral) unit with different kinetic (and exchange) energies. 
However, in the case of the checkerboard lattice, translation symmetry is 
broken explicitly, although once again there is no charge order.
This symmetry-breaking occurs because of perfect nesting of the square Fermi surface~\cite{Indergand2}. 
Alternatively, it can be physically understood as special, local (resonant) states 
formed on the building units (the crossed plaquettes) when the filling is such that these are preferred; 
the occupation states of the units can also be considered to differ.
This local picture is in fact fully equivalent to the nesting instability of the Fermi surface.

\subsection{Metal-insulator transitions and frustrated charge order}
\label{ssec:3b}

We consider next the strong-coupling limit, where at $U = \infty$ one 
obtains the Hamiltonian
\begin{eqnarray}
\label{Hstrong}
H_{\tJV} & = & \mathcal{P} H_0 \mathcal{P} + J \sum_{\langle ij \rangle}
{\bf S}_i \cdot {\bf S}_j + V \sum_{\langle ij \rangle} n_i n_j,\\
\label{HstrondED}
& = & H_{\tJ} + V'\sum_{\langle ij \rangle} n_i n_j,
\end{eqnarray}
where $\mathcal{P}$ is the projection operator enforcing the single-occupancy 
constraint and $V' = V + J/4$. For $V' = 0$, the strong-coupling Hamiltonian 
reduces to the conventional $\tJ$ model.

For simplicity, we restrict our considerations to the checkerboard lattice 
and state only that similar behavior can be found for the kagome lattice.
Examining first the limit $V = \infty$ for the special, commensurate filling
$n = 1/4$ (1/8-filling), the minimum ``classical'' energy ($E = 0$) is 
obtained for all configurations fulfilling the ``ice rule'' of precisely 
one particle on every tetrahedron, as in Fig.~\ref{mapping_QDM}(b). The 
full Hilbert space is then obtained from all possible ways of ``decorating'' 
all dimers with a spin index, {\it i.e.}~the Hilbert space at $n = 1/4$
is exactly that of a two-color QDM~\cite{note_two_color}. A similar procedure, 
decorating the simple loop configurations of Fig.~\ref{mapping_QDM}(a), can 
also be employed to construct the two-color loop configurations which 
constitute the constrained Hilbert space at quarter-filling ($n = 1/2$) 
in the large-$V$ limit. It is then clear that, for these special fillings, 
the GS should be insulating at sufficiently large $V$. The effective dimer 
(or loop) dynamics can be obtained by perturbation in $t/V$. We note that 
in the original derivation~\cite{Pollmann}, for spinless fermions, the 
lowest-order processes were of third order, whereas when spin degrees of 
freedom are included, terms of dimer-exchange type (below) arise at 
second order in $t$. Although the constrained quantum dynamics of fermions 
without~\cite{Pollmann} and with~\cite{Checkerboard_1,Checkerboard_2} spin 
differ, the phase diagrams of these models contain a rich variety of
crystalline phases, breaking lattice translational and/or rotational 
symmetry, in both cases. We postpone to Sec.~\ref{ssec:3c} a discussion of 
the properties of this type of system away from commensurate filling, and 
remark only that, among the novel phenomena which can arise, one of the more 
exotic is the fractionalization under some conditions of a single doped 
charge $e$ into two $e/2$ components~\cite{fractional,Pollmann,note1}.

\begin{figure}
\centerline{\includegraphics*[angle=0,width=0.9\linewidth]{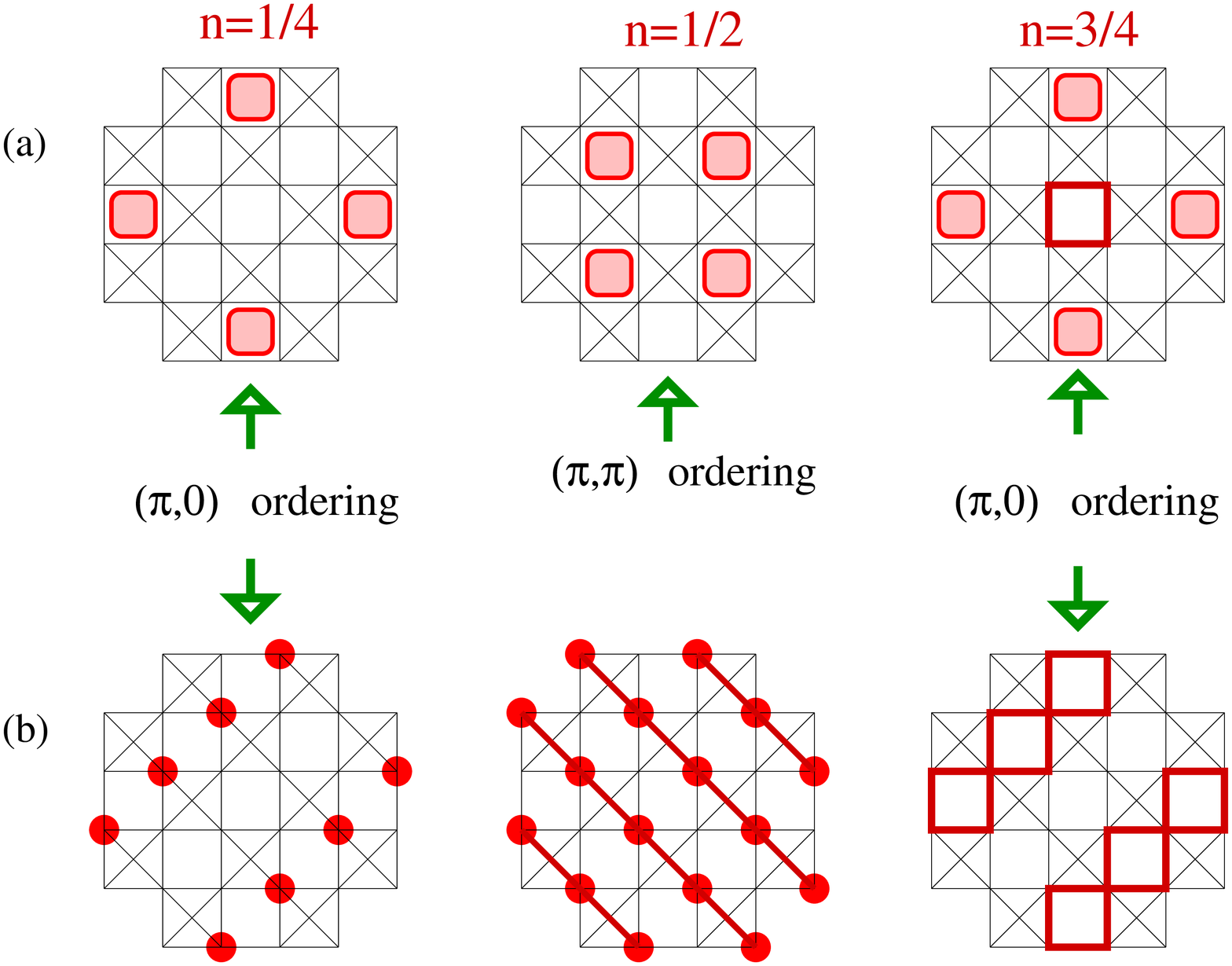}}
\caption{\label{fig:phases} Schematic representation of candidate 
plaquette (a) and columnar (b) phases of the checkerboard lattice for 
electron densities $n = 1/4$, $1/2$, and $3/4$, as discussed in subsection 
\protect\ref{ssec:3b}. Dots, shaded plaquettes, and thick (red) lines 
correspond respectively to electrons, singlet pairs resonating on a 
plaquette, and resonating four-electron plaquette singlets. From 
Ref.~\protect\cite{Checkerboard_2}.}
\end{figure}

Let us now focus in more detail on the insulating phases and consider the 
effective Hamiltonian acting within the constrained Hilbert space as 
second-order processes preserving the ice rule. Here 
${\tilde{\mathcal H}} = H_{\usebox{\mysquare}} + H_J$ with
\begin{eqnarray}
\label{eq:t2}
H_{\usebox{\mysquare}} & = & - t_2
 \sum_{s} P_{\usebox{\mysquare}} (s) ,
\\
 P_{\usebox{\mysquare}} (s) & = & \left(
     c_{i\uparrow}^\dagger c_{j\downarrow}^\dagger
    -c_{i\downarrow}^\dagger c_{j\uparrow}^\dagger
  \right)
  \left(
     c_{k\downarrow}^{\phantom{\dagger}}
     c_{l\uparrow}^{\phantom{\dagger}}
    -c_{k\uparrow}^{\phantom{\dagger}}
     c_{l\downarrow}^{\phantom{\dagger}}
  \right) \\
  & & + \left(
     c_{k\uparrow}^\dagger c_{l\downarrow}^\dagger
    -c_{k\downarrow}^\dagger c_{l\uparrow}^\dagger
  \right)
  \left(
     c_{i\downarrow}^{\phantom{\dagger}}
     c_{j\uparrow}^{\phantom{\dagger}}
    -c_{i\uparrow}^{\phantom{\dagger}}
     c_{j\downarrow}^{\phantom{\dagger}}
  \right),
  \nonumber
\end{eqnarray}
where $t_2 = \frac{2t^2}{V}$ and the index $s$ labels the empty plaquettes 
of the checkerboard lattice (Fig.~\ref{fig:Lattices}); the sites of a 
plaquette $s$ are ordered as $i,k,j,l$ in a clockwise (or anti-clockwise) 
direction. The operator $P_{\usebox{\mysquare}} (s)$ acts on two electrons
forming a singlet bond on one of the two diagonals of $s$, to rotate this 
bond by 90 degrees. Candidate GSs breaking the translational symmetry of 
the lattice are shown in Fig.~\ref{fig:phases}. For a quarter-filled band 
($n = 1/2$), the Resonating-Singlet-Pair Crystal (RSPC) of 
Fig.~\ref{fig:phases}(a) was shown to be stable for $J/t_2 < 
1$~\cite{Checkerboard_1,Checkerboard_2}. In the twofold-degenerate RSPC, 
electron pairs resonate on every second empty plaquette, breaking 
translational symmetry. One therefore expects, on increasing $V/t$ 
and $U/t$, a first-order transition from the BOW state discussed 
above~\cite{Indergand2} to the RSPC. The same analysis also provides 
evidence~\cite{Checkerboard_2} that the system exhibits plaquette order 
of the RSPC type also at $n = 1/4$ or $n = 3/4$, albeit with a quadrupling 
of the lattice unit cell (as opposed to the doubling found for $n = 1/2$) 
and a fourfold-degenerate GS. Qualitative differences between these models 
and their bosonic analogs, which are known for example to exhibit columnar 
order at $n = 1/4$~\cite{QLM}, emphasize the important role of the spin
degrees of freedom, not least in stabilizing plaquette phases over phases 
breaking the rotational symmetry. However, the possibility is being 
investigated~\cite{RSPC_Trousselet} that mixed columnar-plaquette 
phases, similar to one discovered recently in the square-lattice 
QDM~\cite{Ralko_2007_2} and which break both $\pi/2$ rotational
symmetry (as does the columnar phase) and translational symmetry in two 
perpendicular directions (as does the plaquette phase), could be stable 
in some simple and natural extensions of the Hamiltonian~(\ref{eq:t2}).

\subsection{Away from commensurability: doping the Resonating-Singlet-Pair 
Crystal}
\label{ssec:3c}

Whether plaquette ordering of the RSPC type can survive at sufficiently 
low but finite hole (electron) dopant concentrations $x = 1/2 - n$ ($x = 
n - 1/2$) remains unsettled. It has been shown~\cite{Checkerboard_1} that 
phase separation, a generic feature of correlated systems in the vicinity 
of a Mott phase, is restricted to low hole kinetic energies, meaning to 
small $t/J$ and $t/t_2$ ratios, which leaves an extended regime over which 
unconventional superconducting pairing may occur. However, the phase diagram 
of these model for arbitrary electron densities remains largely unexplored, 
and can be expected to harbor further surprises.

\section{Summary}
 
 The 
richness and diversity of systems of doped, frustrated magnets and of itinerant correlated electrons on frustrated 
lattices have been illustrated on selected didactic examples. 
 The dynamics of a small number of doped holes has been investigated in the 
2D kagome and checkerboard Heisenberg quantum 
antiferromagnets revealing striking differences attributed to the different nature of their
non-magnetic GS.
 Two classes of (weakly) doped 
QDMs have also been discussed, which differ in the assumed statistics, bosonic or 
fermionic, of the bare holes. We have proceeded further by considering strongly 
correlated electrons on frustrated triangular lattices, and discuss the 
physics of an unconventional, reentrant metal-insulator transition.
Assuming that increasing pressure reduce 
on-site correlations, the calculated phase diagram reproduces qualitatively the 
essential features of the phase diagram of the molecular solid $\kappa$-(ET)$_2$Cu[N(CN)$_2$]Cl.  
As our final example, we consider correlated fermions moving on frustrated 
lattices at special, commensurate densities for which exotic 
insulating ground states (GSs) are stabilized. 
Interesting similarities with frustrated Heisenberg magnets showing an extensive degeneracy of the 
classical GS manifold are outlined and are shown to be at the heart of their fascinating properties.


{\bf Acknowledgements:} D.P. thanks the French Research Council (ANR) for
support under Grant ANR-05-BLAN-0043-01 and IDRIS (Orsay, France) for allocation of CPU-time
on supercomputers.
H.T. thanks his collaborators Takuma Ohashi, Norio Kawakami,
and Tsutomu Momoi for invaluable discussions.
He also acknowledges support by 
Grants-in-Aid for Scientific Research (Nos.~17071011 and 
19052003), and also by The Next-Generation Supercomputing Project,  
Nanoscience Program, MEXT of Japan.

\printindex


\begin{thebibliography}{99.}
\bibitem{note_two_color} For $n = 3/4$, empty bonds would be decorated if 
the dimers are considered as equivalent to the absence of electrons.
\bibitem{note1} In general, the two components are bound
by a shallow ``string'' potential.
\end{thebibliography}
\end{document}